\documentclass{elsart1p} 
\usepackage{graphicx} 
\usepackage{amssymb} 

\newcommand{\ee}{e^{+} e^{-}}

\newcommand{\ccbar}{c\bar{c}}

\newcommand{\jp}{J/\psi}
\newcommand{\psip}{\psi '}

\newcommand{\pipi}{\pi^{+}\pi^{-}}

\newcommand{\rt}{\rightarrow}

\begin{document} 

\begin{frontmatter} 

\title{Hadronic Spectrum - Multiquark States} 

\author{Stephen L. Olsen\ead{solsen@phys.hawaii.edu}} 
\address{Department of Physics \& Astronomy, University of Hawaii,
Honolulu, HI 96822, USA}

\begin{abstract} 
Many newly discovered mesons behave like $\ccbar$ charmonium states
in that they preferentially decay into final states that contain a $c$-
and a $\bar{c}$-quark, but do not fit expectations for any of the
unfilled levels of the conventional $\ccbar$ spectrum.  There is a
growing  suspicion that at least some of these states are {\it exotic},
{\it i.e.} have a substructure that is more complex than the
quark-antiquark mesons of the classical constituent quark model.
Some of these candidate states have a non-zero electric charge and, thus,
a minimal quark content of $\ccbar u\bar{d}$ or $\ccbar d\bar{u}$.
In addition, states with similar properties have been observed
in the $b$- and $s$-quark sectors.  In this report, the experimental
situation  is briefly reviewed.
\end{abstract} 


\end{frontmatter}

\section{Introduction} 

Quantum Chromodynamics (QCD) suggests the possible
existence of hadrons with a substructure that is more complex
than the three quark baryons and the quark-antiquark mesons
of the Quark Parton Model (QPM).  Possibilities for these
so-called {\it exotic} hadrons include
pentaquark baryons ($qqq\bar{q}q$), tetraquark mesons
($q\bar{q}q\bar{q}$) and quark-gluon hybrids ($q\bar{q}g$).
Although considerable theoretical and experimental effort 
has gone into identifying exotic states, the situation 
remains unclear.  The interest in this subject is demonstrated
by the huge literature related to the purported observation
of the $\Theta(1535)$ strangeness=+1 pentaquark.  According to  
SPIRES, the experimental paper~\cite{nakano_theta}  
that claimed first observation of the $\Theta(1530)$
has received over 785 citations.

There has been some recent progress in the identification 
of what may be exotic mesons.  The BaBar and Belle
$B$-factory experiments have, somewhat unexpectedly,
discovered a number of interesting charmonium-like meson states
that have defied assignment to any of the unfilled levels
of the $\ccbar$ spectrum and, thus, remain unclassified.
These have come to be known
collectively as the ``$XYZ$'' mesons, and
include: the experimentally well established
$X(3872)$~\cite{belle_x3872} and
 $Y(4260)$~\cite{babar_y4260}, which decay to
$\pipi\jp$; the $X(3940)$~\cite{belle_x3940}, seen 
in $D^*\bar{D}$~\cite{charge_conjugate},
and the $X(4160)$~\cite{belle_x4160}
seen in $D^*\bar{D}^*$;  the
$Y(3940)$~\cite{belle_y3940,babar_y3940}, seen in $\omega\jp$;
and the $Y(4350)$~\cite{babar_y4325} 
and $Y(4660)$~\cite{belle_y4350} seen in
$\pi^+\pi^- \psip$.  In addition, Belle reported observations of similar
states but with non-zero electric charge:  the
$Z(4430)$~\cite{belle_z4430}
seen in $\pi^+\psip$ and the $Z_1(4040)$ \&
$Z_2(4240)$~\cite{belle_z4040} seen in $\pi^+\chi_{c1}$.  These
$Z$ states have not yet been confirmed by other
experiments and remain somewhat controversial~\cite{babar_z4430}.
Table~{\ref{table1}} summarizes the
abovementioned $XYZ$ candidate states
as well as some other states discussed below.

In this report I will briefly review the reasons why these states
have eluded conventional $\ccbar$ assignments, discuss possible
alternative interpretations, and present some evidence for similar 
states in the $s$- and $b$-quark sectors.

\begin{table}[htb!]
\begin{center}
\caption{\label{table1} Summary of the candidate $XYZ$ mesons discussed
in this talk.}
\footnotesize
\begin{tabular*}{180mm}{lcccccc}
\hline\hline
state    & $M$~(MeV) &$\Gamma$~(MeV)    & $J^{PC}$ & Decay Modes        &
Production  Modes & Observed by:\\\hline
$Y_s(2175)$& $2175\pm8$&$ 58\pm26 $& $1^{--}$ & $\phi f_0(980)$     &
$\ee$~(ISR), $\jp\rt\eta Y_s(2175)$  &  BaBar, BESII, Belle\\
$X(3872)$& $3871.4\pm0.6$&$<2.3$& $1^{++}$ & $\pipi
\jp$,$\gamma \jp$,
$D\bar{D^*}$
 & $B\rt KX(3872)$, $p\bar{p}$ & Belle, CDF, D0, BaBar\\
$Z(3930)$& $3929\pm5$&$ 29\pm10 $& $2^{++}$ & $D\bar{D}$   &
$\gamma\gamma\rt Z(3940)$ & Belle \\
$X(3940)$& $3942\pm9$&$ 37\pm17 $& $0^{?+}$ & $D\bar{D^*}$ (not
$D\bar{D}$ or $\omega J/\psi$) & $\ee\rt \jp  X(3940)$ & Belle  \\
$Y(3940)$& $3943\pm17$&$ 87\pm34 $&$?^{?+}$ & $\omega J/\psi$ (not
$D\bar{D^*}$) & $B\rt K Y(3940)$ & Belle, BaBar \\
$Y(4008)$& $4008^{+82}_{-49}$&$ 226^{+97}_{-80}$ &$1^{--}$& $\pipi \jp$ &
$\ee$(ISR) & Belle \\
$X(4160)$& $4156\pm29$&$ 139^{+113}_{-65}$ &$0^{?+}$& $D^*\bar{D^*}$
 (not $D\bar{D}$) & $\ee \rt \jp X(4160)$  & Belle \\
$Y(4260)$& $4264\pm12$&$ 83\pm22$ &$1^{--}$&  $\pipi \jp$ & $\ee$(ISR)
&BaBar, CLEO, Belle    \\
$Y(4350)$& $4361\pm13$&$ 74\pm18$ &$1^{--}$&  $\pipi \psip$ & $\ee$(ISR)
& BaBar, Belle  \\
$Y(4660)$& $4664\pm12$&$ 48\pm15 $ &$1^{--}$&  $\pipi \psip$ & $\ee $(ISR)
& Belle     \\
$Z_1(4050)$& $4051^{+24}_{-23}$&$ 82^{+51}_{-29}$ & ? &
$\pi^{\pm}\chi_{c1}$ & $B\rt K Z_1^{\pm}(4050)$  & Belle \\
$Z_2(4250)$& $4248^{+185}_{-45}$&$ 177^{+320}_{-72}$ & ? &
$\pi^{\pm}\chi_{c1}$ & $B\rt K
Z_2^{\pm}(4250)$  & Belle \\
$Z(4430)$& $4433\pm5$&$ 45^{+35}_{-18}$ & ? & $\pi^{\pm}\psip$ & $B\rt K
Z^{\pm}(4430)$  & Belle \\
$Y_b(10890)$  & $10,890\pm 3$ & $55\pm 9$ &   $1^{--}$ &
$\pipi\Upsilon(1,2,3S)$
& $\ee\rt Y_b$ & Belle \\
\hline\hline
\end{tabular*}%
\end{center}
\end{table}

\section{Charmonium possibilities}
\label{charmonium}

\begin{figure}[t] 
\includegraphics[scale=0.43]{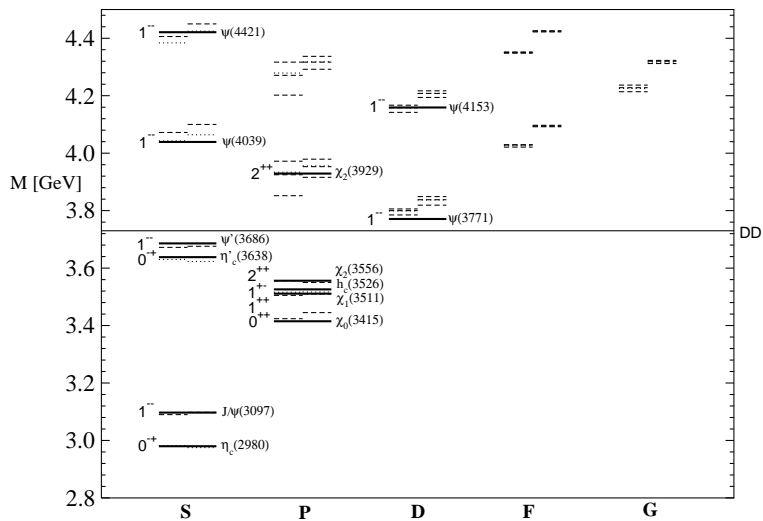}
\caption{The predicted and observed spectrum of $\ccbar$ charmonium 
mesons.  Already assigned states and their experimentally 
measured masses are indicated by solid bars and their 
commonly used names.  The broken lines indicate
various theoretical predictions.  The horizontal line
at 3.73~GeV indicates the mass threshold for decays
to $D\bar{D}$ ``open charm'' final states.}
\label{fig:charmonium_spectrum}
\end{figure}

The $\ccbar$ charmonium meson level diagram is shown in
Fig.~\ref{fig:charmonium_spectrum}~\cite{charmonium_spectrum}. Here the
states that have already been assigned are labeled by their commonly used
symbols and measured mass values.  The solid lines indicate the 
measured levels and the broken lines indicate masses derived from 
QCD-motivated potential model calculations~\cite{BGS}.  If any of the 
$XYZ$ mesons are to be interpreted as simple quark-antiquark states, 
they must be assigned to one of the figure's unlabeled levels.
All of the states with mass below the $M=2m_D=3.73$~GeV 
``open-charm'' threshold (indicated by the horizontal line in
Fig.~\ref{fig:charmonium_spectrum}) have already been identified and
have properties that are in good agreement with potential model
expectations.  In addition, all of the $1^{--}$ levels above the
open charm threshold have been assigned to peaks in
the total annihilation cross section for 
$\ee\rt~hadrons$~\cite{bes_R-fit}.

Of the $XYZ$ states listed in Table~\ref{table1}, only the
$Z(3930)$~\cite{belle_z3930} has been convincingly assigned to a 
charmonium 
level;
there is general agreement that this is the ($2^3P_2$)
$\chi_{c2}^\prime$.

\subsection{The $X(3872)$}

The experimentally preferred $J^{PC}$ value for the $X(3872)$
is $1^{++}$, although $2^{-+}$ has not been conclusively ruled
out~\cite{CDF_jpc}.  The only unfilled $1^{++}$ level in
Fig.~\ref{fig:charmonium_spectrum} is the $\chi_{c1}^\prime$, the
$2^3P_1$ $\ccbar$ state.   As mentioned above, the $J=2$
triplet partner state
for this level has been identified as the $Z(3930)$
with a mass of $3929\pm 5$~MeV.  A
$\chi_{c1}^\prime$ assignment for the $X(3872)$ would imply
a $\chi_{c2}^{\prime}$-$\chi_{c1}^{\prime}$ mass splitting for 
radial quantum number
$n=2$ ({\it i.e.} $\delta m \sim 57$~MeV) that is larger than that
for the $n=1$ splitting ($\delta m = 46$~MeV), contrary to 
potential model expectations.  A bigger difficulty with this 
assignment is the fact that the
$X(3872)\rt\rho\jp$ discovery mode would be an isospin violating
transition that should be strongly suppressed compared to
the  $X(3872)\rt \gamma\jp$ mode; the latter is measured to
be much smaller than the former~\cite{x3872_2gammajpsi}.
Using the $2^{-+}$ assignment does not help; in this
case the $\pipi\jp$ mode would also be isospin violating, and the
$\gamma\jp$ transition, which would be a $\Delta L=2$ transition,
would be unmeasurably small, which it isn't.

\subsection{The $X(3940)$ and $X(4160)$}

The $X(3940)$ is seen in the $D\bar{D^*}$ system recoiling from
the $\jp$ in exclusive $ee\rt\jp D\bar{D^*}$ annihilations;
the $X(4160)$ is seen in the $D^*\bar{D^*}$ system in
$\jp$ in $ee\rt\jp D^*\bar{D^*}$.  Neither are seen in the
experimentally more accessible $D\bar{D}$ channel.  The
only known charmonium states that are seen recoiling from the $\jp$
in $\ee\rt\jp X$ processes have $J=0$. This, plus the
absence of the $D\bar{D}$ mode, 
provides circumstantial evidence that favors $J^{PC}=0^{-+}$ assignments
for both states, which for charmonium would be the $\eta_c^{\prime\prime}$
and $\eta_c^{\prime\prime\prime}$.  Such an assignment has difficulty
with the measured masses: the predicted $\eta_c^{\prime\prime}$ mass
is about 4050~MeV, over 100~MeV too high for the $X(3940)$; the
predicted $\eta_c^{\prime\prime\prime}$ is around 4400~MeV, more
than 200~MeV higher than the $X(4160)$,

\subsection{The $Y(3940)$}

The $Y(3940)$ was first seen by Belle
as a near-threshold peak in the $\omega\jp$
invariant mass spectrum in exclusive $B\rt K\omega\jp$
decays~\cite{belle_y3940}.  It was subsequently confirmed
by BaBar~\cite{babar_y3940}, although there remain some 
($\sim 2\sigma$) discrepancies between the Belle \& BaBar 
measurements  of the mass and width.  

It is unlikely
that the $Y(3940)$ (seen in $\omega\jp$)  and the 
$X(3940)$ (seen in $D\bar{D^*}$) are different
decay modes of the same state.  Belle has searched for
$Y(3940)\rt D\bar{D^*}$ in $B\rt K D\bar{D^*}$
decays and finds a 90\%~CL lower limit of 
$\mathcal{B}(Y\rt\omega\jp)/
\mathcal{B}(Y\rt D\bar{D^*})>0.71$~\cite{belle_y3940_ddstr}
that contradicts a 90\%~CL upper limit from a search for
$X(3940)\rt\omega\jp$ in $\ee\rt \jp\jp\omega$ annihilations:
$\mathcal{B}(X\rt\omega\jp)/
\mathcal{B}(X\rt D\bar{D^*})<0.58$~\cite{belle_x3940}.

Possible charmonium assignments for the $Y(3940)$ are the
$\eta_c^{\prime\prime}$ ($0^{-+}$) --- although its mass
is a little low --- and the $\chi_{c0}^{\prime}$ ($0^{++}$) 
for which its mass is too high.   The primary
difficulty with a charmonium assignment for the $Y(3940)$ is
its large partial width to $\omega\jp$, which reasonable
estimates put above 1~MeV~\cite{godfrey_olsen}
and which may in fact be quite a bit higher.  This is
well above the measured partial widths for any of the
observed hadronic transitions between charmonium states.

\subsection{The $J^{PC}=1^{--}$ $Y$ states}

The $Y(4260)$ was first seen by BaBar as a peak in the 
$\pipi\jp$ mass spectrum in the initial-state-radiation 
(ISR)  process $\ee\rt\gamma_{ISR}\pipi\jp$.
The $Y(4350)$ and $Y(4660)$ are seen in the $\pipi\psip$ mass
spectrum in the $ee\rt\pipi\psip$ ISR process.  (Belle also sees
a broad $Y(4008)$ peak in $\pipi\jp$~\cite{belle_y4008}, but this
has not been confirmed by BaBar~\cite{babar_y4008}.)  Since these
states are produced via ISR, their $J^{PC}$ has to be $1^{--}$.

There are no unassigned $1^{--}$ slots for any of these states in 
the $M<4.4$~GeV spectrum of Fig.~\ref{fig:charmonium_spectrum}.
Moreover, no hint of any of them in seen in any of the
$D^{(*)}\bar{D^{(*)}}$ channels~\cite{galina}. This implies
that their $\pipi\jp(\psip)$ decay widths must be quite large.  
In the case of the $Y(4260)$, the $\pipi\jp$ has been
established to be more that 1.6~MeV~\cite{mo_xiaohu}. This is much too 
large for charmonium, where allowed $\pipi\jp$ transitions have
measured partial widths of 100~keV or less.

\subsection{The charged $Z$ particles}

\begin{figure}[h]
\begin{minipage}[t]{75mm}
\includegraphics[scale=0.40]{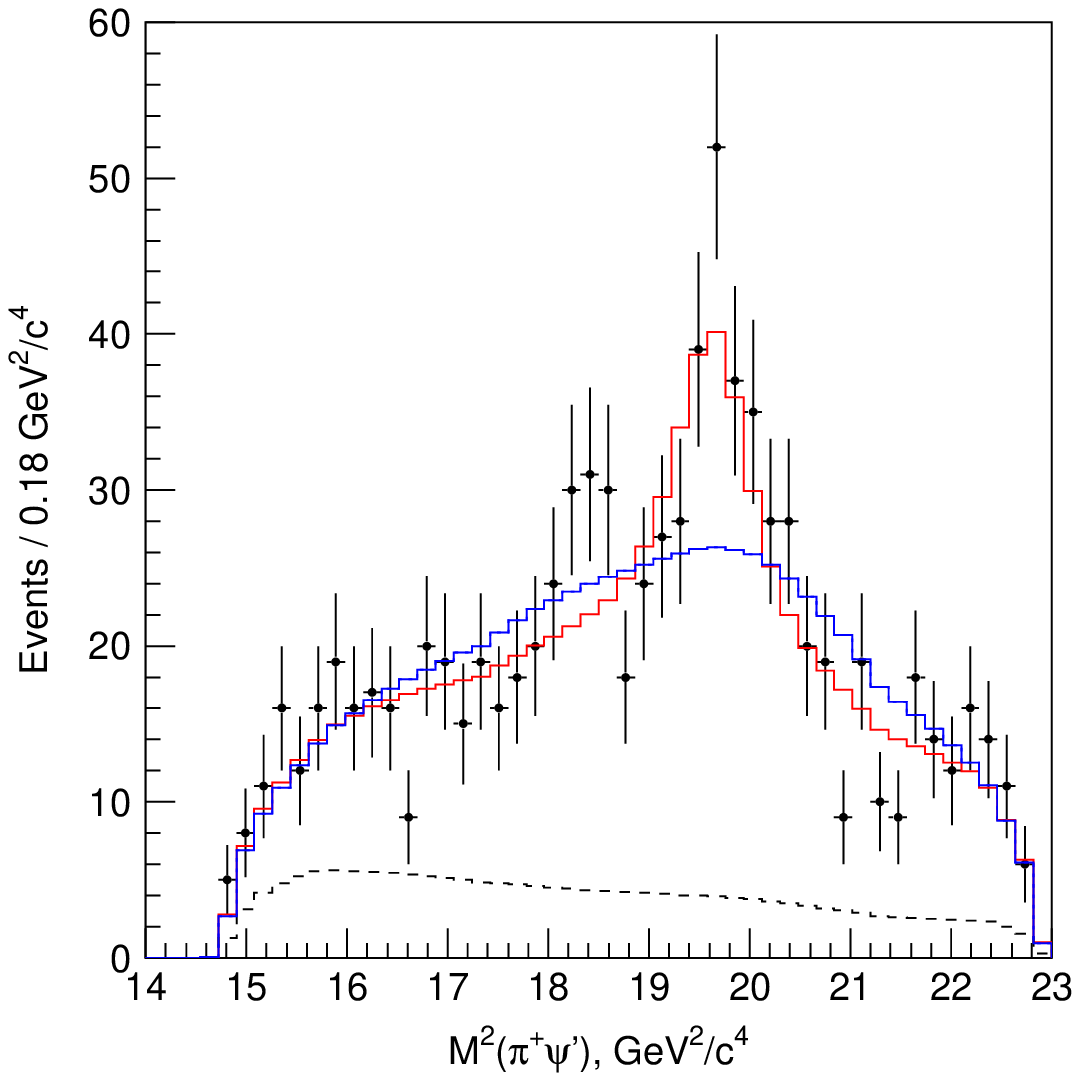}
\end{minipage}
\hspace{\fill}
\begin{minipage}[t]{75mm}
\includegraphics[scale=0.55]{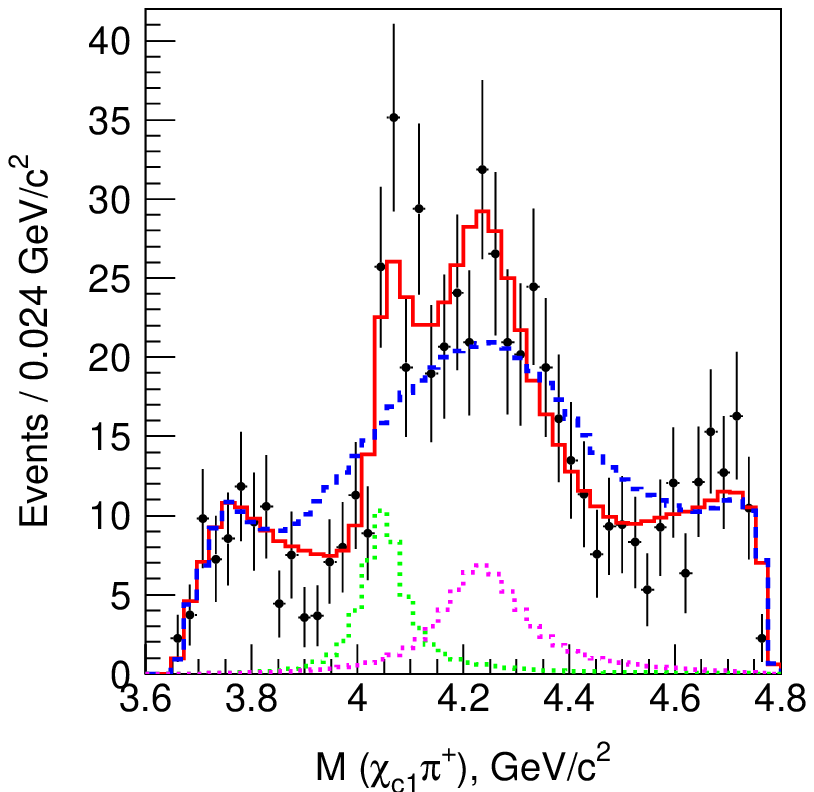}
\end{minipage}
\caption{\label{fig:z_projections} {\bf Left:} ($B\rt K\pi^{\pm}\psip$)
The points with errors are data and the histograms show fits 
to a Dalitz-plot projection that has the $K^*$ bands removed with \& 
without the inclusion of the $Z(4430)$ resonance in the 
$\pi^{\pm}\psip$ channel.
{\bf Right:} ($B\rt K\pi^{\pm}\chi_{c1}$)
A similar Dalitz-plot projection for data \& fits with \& without
the inclusion of two resonances in the $\pi^{\pm}\chi_{c1}$ channel.
}
\end{figure}

Belle reported a peak with a $\sim 6.5\sigma$ statistical significance 
near 4430~MeV in  the $\pi^{\pm}\psip$ channel
in exclusive  $B\rt K\pi^{\pm}\psip$ decays~\cite{belle_z4430}.
A peak at the observed $\pi\psip$ invariant mass value
cannot be produced by reflections from the $K\pi$ system.
However, BaBar did not confirm this peak, finding at most a
signal of $\sim 1.7\sigma$ significance~\cite{babar_z4430}.
A subsequent full Dalitz-plot analysis 
(see Fig.~\ref{fig:z_projections}  left)
of the Belle $B\rt 
K\pi^{\pm}\psip$ sample confirms their
original mass and significance determinations~\cite{chistov_qwg}.   
 Belle also reported two peaks with greater than $5\sigma$
statistical significance in the $\pi^{\pm}\chi_{c1}$ channel,
the $Z_1(4050)$ \& $Z_2(4250)$, in exclusive
$B\rt K\pi^{\pm}\chi_{c1}$ decays, again from a  Dalitz-plot
analysis (Fig.~\ref{fig:z_projections} right). 
If these peaks are interpreted as meson states, they must have a 
minimal tetraquark $\ccbar u\bar{d}$  
substructure and there are no possible charmonium 
or charmonium hybrid assignments.

\section{Exotic possibilities}
\label{exotic}
In this section I discuss the possible interpretations of the
$XYZ$ peaks as $c\bar{c}q\bar{q}$ tetraquark states or $c\bar{c}$-gluon 
hybrid states.

\subsection{Tetraquarks}
Two very distinct types of tetraquark mesons have been proposed:
{\it molecular states}, which are relatively
loosely bound structures comprised of deuteron-like 
mesons-antimeson bound states~\cite{molecules}, 
and {\it diquark-diantiquark mesons} in which the two quarks form
an anticolor triplet state that binds tightly to a
color triplet that is formed from the two antiquarks~\cite{maiani}.  These
two types of structures have very different phenomenologies.

\subsubsection{Molecules}
A molecular state is expected to have a mass that is
slightly below  the sum of the masses of its 
meson-antimeson constituents and exhibit large
isospin violations.  The $X(3872)$, with a mass
that is within errors of the $m_D + m_{D^*}$ mass
threshold and has decay rates to $\pipi\jp$ and
$\pipi\pi^0\jp$ that are nearly equal~\cite{belle_x23pijpsi}.
Thus, this is a nearly ideal candidate for a $D\bar{D^*}$ molecular 
state, either real~\cite{tornqvist2004,swanson,voloshin,braaten}
or virtual~\cite{hanhart}.  On the other hand, its proximity 
to the $D\bar{D^*}$ threshold has also led to speculation
that it is some kind of a threshold effect~\cite{bugg,chao,zhang}. 
In theses latter schemes, mixing with the $\chi_{c1}^{\prime}$ 
charmonium state can play an important role.

\begin{figure}[h]
\begin{minipage}[t]{75mm}
\includegraphics[scale=0.3]{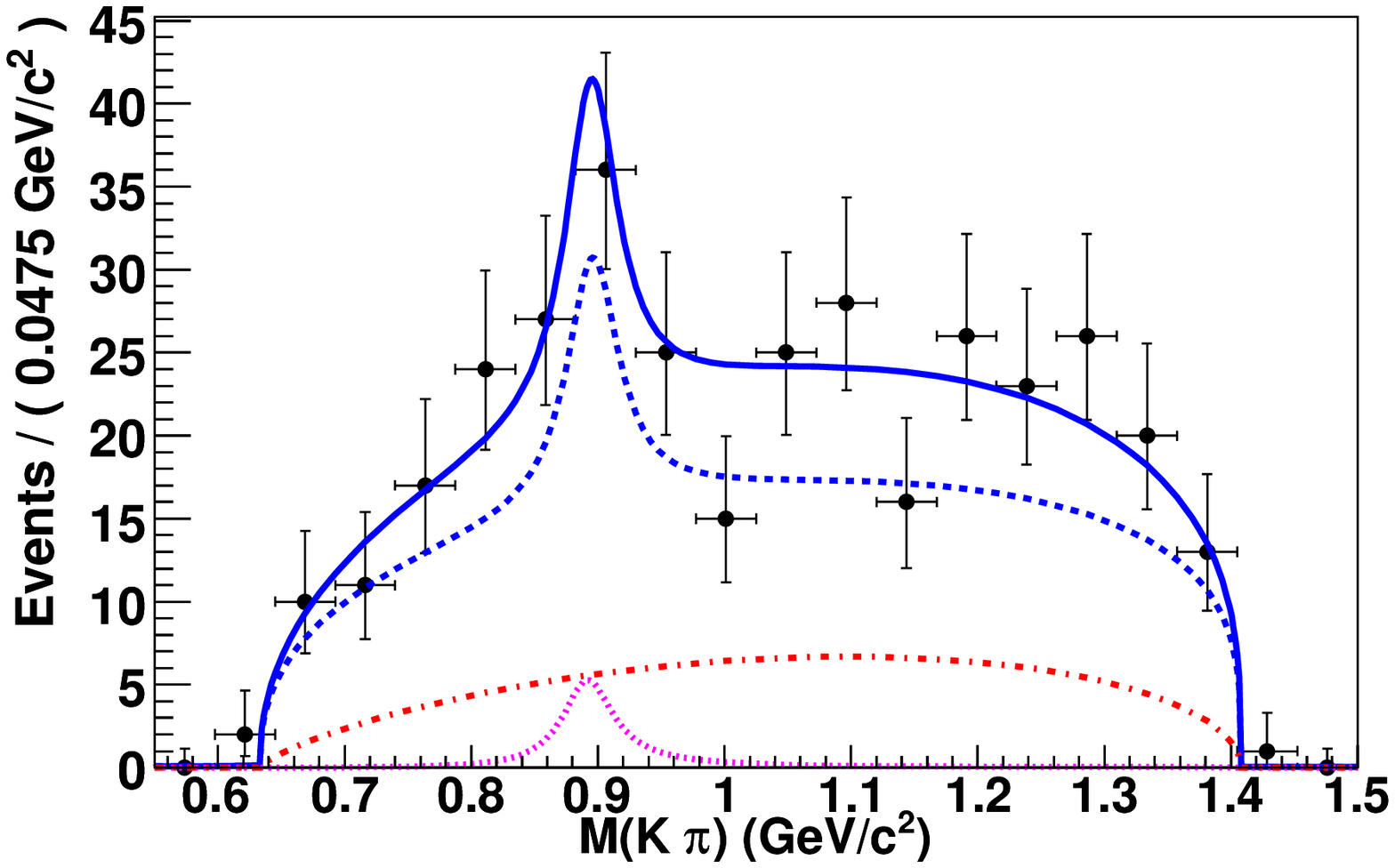}
\end{minipage}
\hspace{\fill}
\begin{minipage}[t]{75mm}
\includegraphics[scale=0.3]{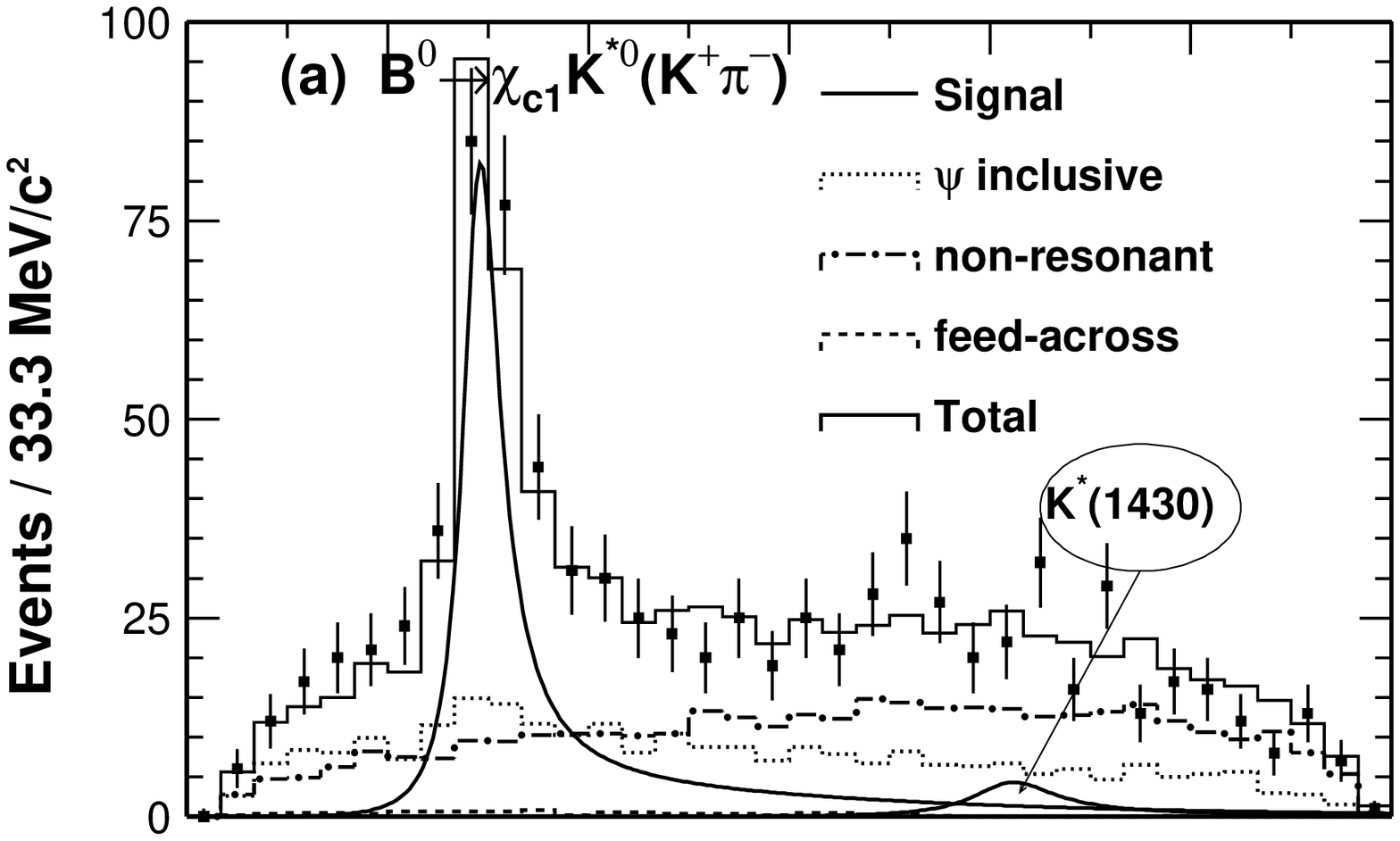}
\end{minipage}
\caption{\label{fig:mkpi} {\bf Left:}
The $M(K\pi)$ distribution for $B\rt K^-\pi^+ X(3872)$ decays.
The lower curves show backgrounds from
non-resonant and $K^*(890)$ $K\pi$ systems. {\bf Right:} 
The $M(K\pi)$ distribution for $B\rt K^-\pi^+ \chi_{c1}$  decays.
}
\end{figure}

A new piece of experimental information, the significance of
which has yet to be commented on by any theorist, is a study
of $X(3872)$ production in exclusive $B\rt K\pi X(3872)$ 
decays~\cite{belle_kpix3872}.
The left panel of Fig.~\ref{fig:mkpi} shows the $K\pi$ invariant
mass distribution, where it is evident that non-resonant
$K\pi$ production dominates, and the $K^*(890)$ contribution
is small and of marginal significance.  This is in contrast to what is 
seen in all other 
$B\rt K\pi$+charmonium decays in which the $K^*(890)$ contribution
dominates; for example,
the right panel of Fig.~\ref{fig:mkpi} shows the
$M(K\pi)$ distribution for $B\rt K^-\pi^+\chi_{c1}$ 
decays~\cite{belle_kpichic1}, where a prominent $K^*(890)$ signal
is clearly evident.  

Not all of the $XYZ$ states fit the molecule picture.
For example the $X(3940)$, $Y(3940)$ and $Y(4660)$ are not
near any $D^{(*)}\bar{D^{(*)}}$ mass threshold. (Note that 
$\pi$-exchange, the dominant binding term in molecular models,
is absent in $D_s^{(*)}\bar{D_s^{(*)}}$ systems.)

\subsubsection{Diquark-diantiquarks}
Essentially all of the observed $XYZ$ states can be accommodated
by the diquark-diantiquark model.    However, in this picture, each
of the assigned state is expected to have an associated flavor-$SU(3)$
multiplet of states.   One prediction of this model is that there
should be two $X(3872)$ states --- $X_u = cu\bar{c}\bar{u}$ and
$X_d = cd\bar{c}\bar{d}$ --- with a mass difference of $8\pm 3$~MeV.
No evidence for such a pairing has been 
found~\cite{belle_y3940_ddstr,belle_kpix3872}.
In addition, in this model one expects
a charged isospin partner of the $X(3872)$ to be produced 
in $B$ decays.  BaBar searched for such an
$X^+(3872)\rt\rho^+\jp$ state in neutral $B$
meson decays and and set an upper limit that is well below
isospin-based expectations~\cite{babar_xplus3872}.  No isospin
partners of any of the other $XYZ$ states have been reported. 

\subsection{Hybrids}
The lattice QCD expectation for the mass of the lowest-lying 
charmonium hybrid is around 4.3~GeV and the relevant open-charm 
threshold is 4.29~GeV, the $D^{**}\bar{D}$ mass threshold, where 
$D^{**}$  denotes  the lowest mass $P$-wave charmed meson with mass 
2.42~GeV.  Since the $Y(4260)$ mass is near the LQCD value and
below the $D^{**}\bar{D}$ mass threshold --- which would explain its
relatively strong decay rate to $\pipi\jp$ as opposed to open
charm states --- a charmonium hybrid interpretation is attractive. 
However, the $Y(4260)$ is broad and $D^{**}\bar{D}$ decays are
accessible from its high mass side, but there is no sign of a lineshape
distortion that might be expected when a dominant new decay
channel opens up.  Moreover, the $Y(4350)$ \& $Y(4660)$ are both
well above all $D^{**}\bar{D}$ thresholds and no open charm decays
have been seen.   So, while the hybrid interpretation might work for
the $Y(4260)$, it does not seem to apply to the other $1^{--}$ 
ISR-produced states.

\section{Evidence for $XYZ$-like states in the $s$- and $b$-quark 
sectors}
\begin{figure}[h]
\begin{minipage}[t]{75mm}
\includegraphics[scale=0.35]{olsen_panic_fig4a.epsi}
\end{minipage}
\hspace{\fill}
\begin{minipage}[t]{75mm}
\includegraphics[scale=0.35]{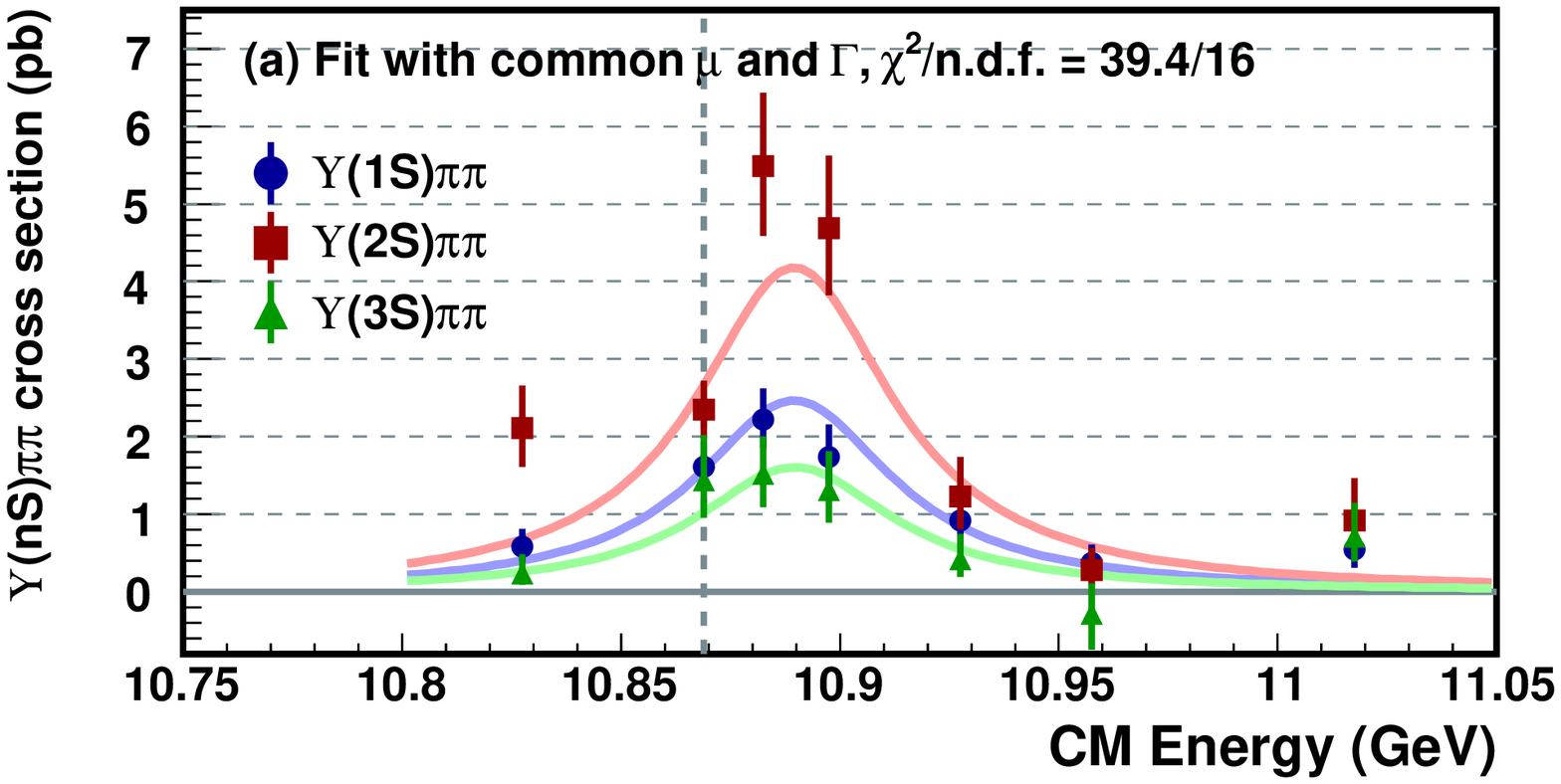}
\end{minipage}
\caption{\label{fig:yb} {\bf Left:}
The $M(f_0(980)\phi))$ distribution for $\ee\rt \gamma_{ISR} f_0\phi$.
{\bf Right:} the energy dependence of $\ee\rt \pipi 
\Upsilon(nS)~(n=1,~2~\&~3)$  
near $\sqrt{s}=10.9$~GeV. In both cases the data are from Belle.}
\end{figure}

An obvious question is whether or not there are counterpart states
in the $s$- and $b$-quark sectors.  Recent results
suggest that there are.  In a BaBar study of the ISR process
$\ee\rt\gamma_{ISR}\pipi\phi$, where the $\pipi$ comes from
$f_0(980)\rt\pipi$, a distinct $\pipi\phi$ mass peak is seen 
at 2175~MeV~\cite{babar_y2175}.  This peak was confirmed in an ISR 
measurement by Belle~\cite{belle_y2175} (left panel of Fig.~\ref{fig:yb})
and seen in $\jp\rt\eta f_0\phi$
decays by BES~\cite{bes_y2175}.  Although a conventional $s\bar{s}$
assignment cannot be ruled out~\cite{mlyan_y2175}, this state has 
properties similar to what one would expect for an $s$-quark sector
counterpart of the $Y(4260)$.

The Belle group recently reported measurements of the energy dependence
of the $\ee\rt\pipi\Upsilon(nS)~(n=1,~2~\&~3)$ cross section around  
$E_{cm}\sim 10.9$~GeV and found peaks in all three channels at 
10.899~GeV (right panel of Fig.~\ref{fig:yb})~\cite{belle_y10899}.  
The peak mass 
and width valuse are quite distinct from those of 
 the nearby $\Upsilon(5S)$ bottomonium ($b\bar{b})$ 
state, and the cross section values are more than two-orders-of-magnitude 
above expectations for a conventional $b\bar{b}$ system.  One
interpretation for this peak is that it is a $b$-quark sector
equivalent of the $1^{--}$ $Y$ states seen in the $c$-quark 
sector~\cite{weishu}.

\section{Summary}
There is a growing body of evidence for a new type of hadron spectroscopy
involving pairs of $c$-quarks that neither fits well to classic Quark 
Parton Model expectations nor QCD-motivated extensions.  A recurring 
feature of these new state are large partial widths for decays to
charmonium plus light hadrons.  There is some evidence for similar 
structures in the $s$- and $b$-quark sectors.

\section{Acknowledgments}
I thank Professor Tserruya and the PANIC organizing committee for their 
gracious hospitality.  I also thank Avraham~Gal
and my colleagues Haibo~Li, Roman~Mizuk,
Chengping~Shen and Bruce~Yabsley
for their assistance in the preparation of this manuscript.


\begin{thebibliography}{00}
\bibitem{charge_conjugate} The inclusion of charge conjugate states is 
always implied.
\end{thebibliography}
\end{document}